\documentclass[journal=nalefd,manuscript=letter,layout=twocolumn]{achemso}

\usepackage{chemformula} 
\usepackage[T1]{fontenc} 

\newcommand{\beginsupplement}{%
        \setcounter{table}{0}
        \renewcommand{\thetable}{S\arabic{table}}%
        \setcounter{figure}{0}
        \renewcommand{\thefigure}{S\arabic{figure}}%
     }

\author{Henrik~Beccard}%
\affiliation{Institute of Applied Physics, Technische Universit\"at Dresden, N\"othnitzer Strasse 61, 01187 Dresden, Germany}%
\author{Benjamin~Kirbus}%
\affiliation{Institute of Applied Physics, Technische Universit\"at Dresden, N\"othnitzer Strasse 61, 01187 Dresden, Germany}%
\author{Elke~Beyreuther}
\email{elke.beyreuther@tu-dresden.de}%
\author{Michael~R\"using}%
\affiliation{Institute of Applied Physics, Technische Universit\"at Dresden, N\"othnitzer Strasse 61, 01187 Dresden, Germany}%
\author{Petr~Bednyakov}
\affiliation{Institute of Physics of the Czech Academy of Sciences, 182~21 Prague~8, Czech Republic}
\author{Jirka~Hlinka}
\affiliation{Institute of Physics of the Czech Academy of Sciences, 182~21 Prague~8, Czech Republic}
\author{Lukas~M.~Eng}
\affiliation{Institute of Applied Physics, Technische Universit\"at Dresden, N\"othnitzer Strasse 61, 01187 Dresden, Germany}%
\altaffiliation{ct.qmat: Dresden-W\"urzburg Cluster of Excellence--EXC 2147, Technische Universit\"at Dresden, 01062 Dresden, Germany}%

\title[]
  {Large Hall electron mobilities in \textit{head-to-head} BaTiO$_3$-domain walls}


\begin{document}







\begin{abstract}
  Strongly charged head-to-head (H2H) domain walls (DWs) that are purposely engineered along the [110] crystallographic orientation into ferroelectric BaTiO$_3$ single crystals have been proposed as novel 2-dimensional electron gases (2DEGs) due to their significant domain wall conductivity (DWC). Here, we quantify these 2DEG properties through dedicated Hall-transport measurements in van-der-Pauw 4-point geometry at room temperature, finding the electron mobility to reach around 400~cm$^2$(Vs)$^{-1}$, while the 2-dimensional charge density amounts to ~7$\times$10$^3$cm$^{-2}$. We underline the necessity to take account of thermal and geometrical-misalignment offset voltages by evaluating the Hall resistance under magnetic-field sweeps, since otherwise dramatic errors of several hundred percent in the derived mobility and charge density values can occur. Apart from the specific characterization of the conducting BaTiO$_3$ DW, we propose the method as an easy and fast way to quantitatively characterize ferroic conducting DWs, complementary to previously proposed scanning-probe-based Hall-potential analyses.
\end{abstract}


\hrulefill
\\~
\hrulefill

Within the vast flurry of interest in various 2-dimensional (2D) electronic systems, which are key elements of contemporary and upcoming nanoelectronic device concepts, domain walls (DWs) in ferroelectrics as potentially versatile functional elements have experienced concerted research activities for the recent decade, which is documented with an increasing number of reviews  \cite{cat12,mei15,slu16,bed18,sha19,nat20,mei21,sha22}. Thereby, especially their possible highly enlarged electrical conductivity as compared to the (mostly) insulating bulk matrix has been in the scientific focus. In early years, the pure proof of the "domain wall conductivity (DWC)" effect itself was reported for a number of ferroelectric crystals and thin films. Subsequently, the ongoing research has been dedicated to the full quantitative and in-depth microscopic understanding of the underlying confined charge carrier transport mechanisms, since a coherent overall picture allows for the development of reproducible protocols to accurately tune this quasi-2D conductivity.

Up to now, the determination of quantitative transport parameters, such as carrier densities and mobilities, has been realized only in a few exemplary cases. In this context, especially the analysis of the Hall-effect of DW-confined charge carriers in the improper ferroelectrics YbMnO$_3$ and ErMnO$_3$ was shown to be an invaluable tool, as reported in the two pioneering works of Campbell et al.\cite{cam16} and Turner et al.\cite{tur18}, respectively. There, the authors choose scanning-probe based approaches, where the contribution of the Hall-potential to the cantilever deflection has to be disentangled in a time-consuming sophisticated manner via calibration routines and accompanied simulations. As a certain constraint, that approach is mainly suitable to extract near-surface charge carrier densities and mobilities, which might, however, differ dramatically from the bulk values due to specific issues such as the occurrence of surface screening charges, band bending, and the complex interface physics of the electrode/DW junction. At this point, our approach comes into play and presents a robust complementary Hall-effect measurement scenario, which is based on adopting the classical van-der-Pauw \cite{vdP58} four-point resistance measurement configuration using macroscopic electrodes to the case of charged ferroic DWs that fully penetrate the crystal. We demonstrate the feasibility of this established and robust method by studying the Hall effect in conductive head-to-head domain walls engineered into [110] BaTiO$_3$ (BTO) single crystals, thus serving as the model system of a DW-based 2-dimensional electron gas (2DEG), where both an orders-of-magnitude enhanced conductivity and a metallic-like temperature dependence of the resistance had been clearly proven\cite{slu13}. The accuracy and integrity of the sheet and volume majority carrier densities and the Hall mobilities, which are extracted from the measured Hall voltages and sheet resistances, are critically discussed, with special care laid on the correction of offset voltages. Notably, the latter may lead to errors of several hundred percents. In contrast to the scanning-probe based methods mentioned above\cite{cam16,tur18}, many of these challenges are well known and have been addressed in the past\cite{wer17b}, which we take advantage of for the goal postulated here.

\begin{figure*}[!ht]
\includegraphics[width=\textwidth]{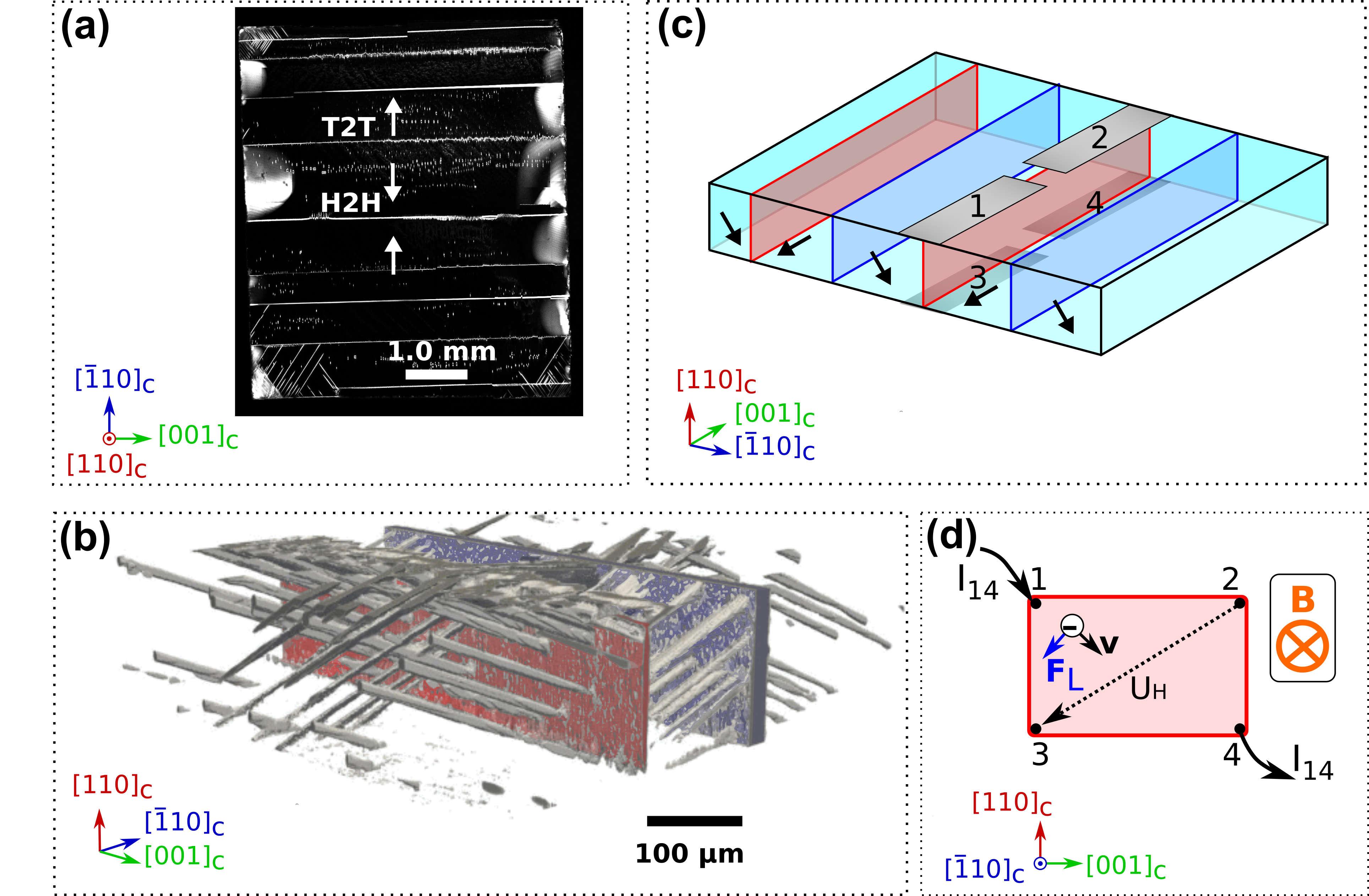}
\caption{\label{fig:1}Geometries of the BaTiO$_3$ [110] domain walls and van-der-Pauw electrode configuration: depicted schematically and visualized by second harmonic generation (SHG) microscopy: (a) 2D top-view SHG image proving two types of 90$\symbol{23}$ domain walls, i.e., T2T and H2H DWs, in (110) BaTiO$_3$. (b) Full 3D rendering of the SHG data, which exhibits a complex domain structure with the red-colored DW representing the conducting H2H type, the blue-colored DW the (non-conducting) T2T type, respectively, while the greyish structures originate from neutral non-conductive DWs aligned in [100] directions. (c) 3D schematic overview of the dipole moment orientation in (110)~BaTiO$_3$ with conductive H2H (red) and non-conductive (blue) T2T domain walls, including also the configuration of chromium electrodes (grey areas with a size of 2.0$\times$0.6~mm$^2$ each), labelled 1--4. (d) Built-up of a Hall voltage within the conductive DW: while the current $I_{14}$ is driven through the DW with a magnetic field $B$ perpendicular to the DW plane, a Lorentz force $F_L$ occurs. The latter deflects the electrons from their original trajectory and thus induces a voltage $U_H$ between electrodes 2 and 3.}
\end{figure*}

In the present study, an undoped (110) BaTiO$_3$ crystal, 5$\times$5$\times$0.5~mm$^3$ in size, underwent a \emph{so-called} frustrative-poling process -- for more details refer to the Supplementary Material -- resulting in planar head-to-head (H2H) and tail-to-tail (T2T) 90$\symbol{23}$~DWs\cite{bed16}, as visible in the second-harmonic generation microscopy images, recorded with a setup described earlier\cite{kir19}, displayed in figs.~\ref{fig:1}(a) and (b), as well as in the schematics of fig.~\ref{fig:1}(c). 

As a precondition for any electrical characterization, four 8-nm-thick chromium electrodes were vapor-deposited under high-vacuum conditions (base pressure: 10$^{-6}$mbar) using a shadow mask. The electrode geometries and positions with respect to the crystal and DW orientations are shown in fig.~\ref{fig:1}(c). In the following, these electrodes are indexed by numbers 1--4, when applying and discussing the van-der-Pauw method.

\begin{figure}[!ht]
\includegraphics[width=0.5\textwidth]{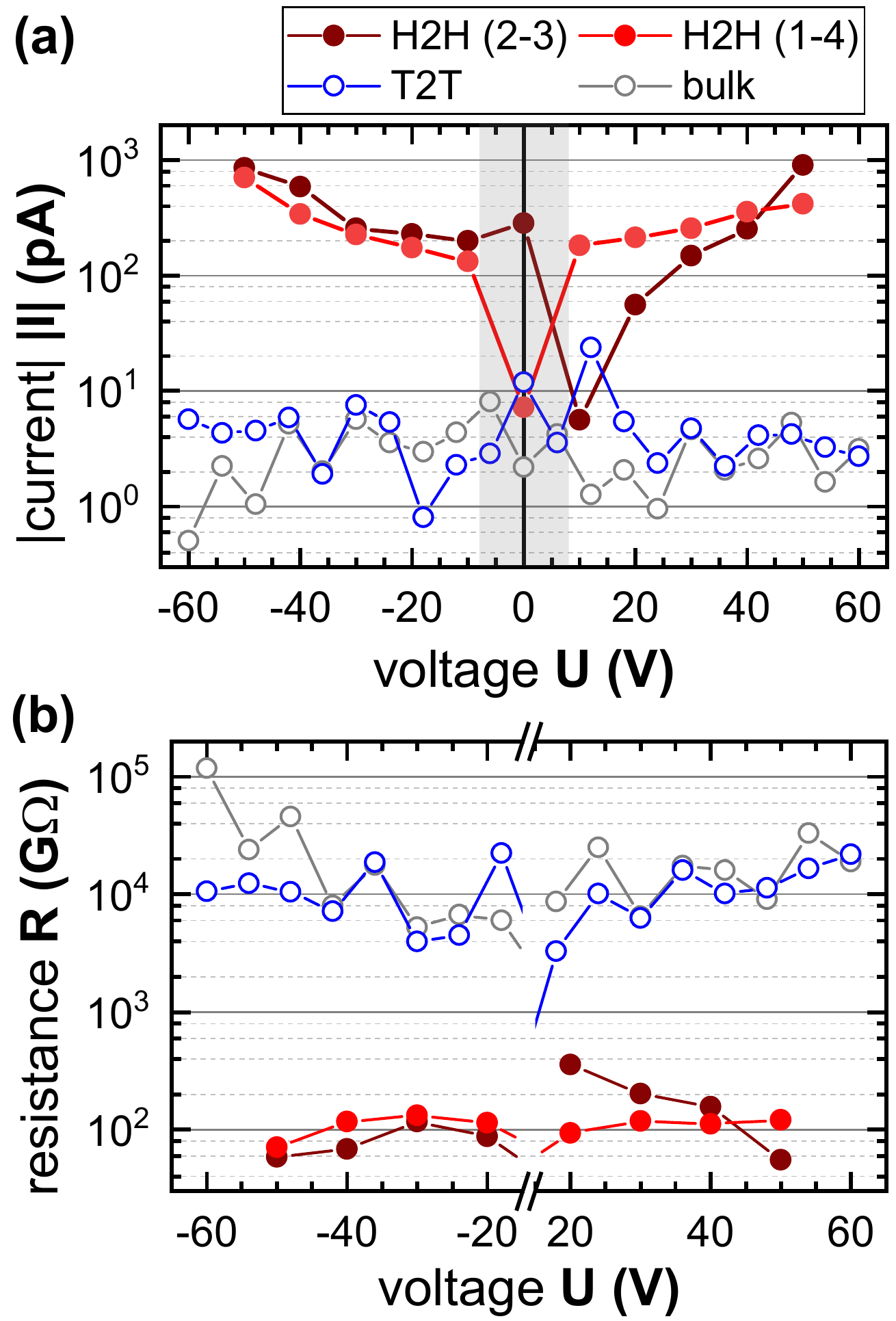}
\caption{\label{fig:2}Comparative fundamental electrical characterization of 90$\symbol{23}$-H2H- and -T2T domain walls as well as the bulk of the (110) BaTiO$_3$ crystal without magnetic field: (a) Absolute value of the current as function of bias voltage plotted on a logarithmic scale. The conductive H2H DW exhibits significantly higher currents as compared to the T2T- and bulk reference cases with partially non-ohmic behavior above certain thresholds, while the T2T DW and the bulk show basically noise. Data points within the greyish range -- which corresponds to the threshold\cite{slu13} of $\pm$8~V, below which the BTO 2DEG is not conductive -- are of limited  validity. (b) Corresponding resistance values as extracted from the IU-curves in panel (a). Additionally, the linear representations of the IU-data can be found in SI-figs.~S1(a) and (c).}
\end{figure}

Prior to any magnetic-field-dependent investigations, the conductivity of the H2H domain wall was investigated by acquiring current-voltage characteristics along both diagonal directions, i.e., between contacts 2-3 and 1-4 in the $\pm$50-V~range [cf. sketch in fig.~\ref{fig:1}(d)]. These results are displayed in fig.~\ref{fig:2}(a) in a semilogarithmic plot and, alternatively, linearly scaled in the SI-fig.~S1(a), showing in principle a similar behavior for both directions apart from a stronger current offset at zero bias voltage for the 2-3~diagonal and a different onset of non-ohmic behavior. Thus the linear fits [straight lines in SI-fig.~S1(a)], which provide values of 239 and 108~G$\Omega$ for $R_{14}$ and $R_{23}$, respectively, are seen as rough estimates only. Alternatively, the (voltage- dependent) resistances are depicted in fig.~\ref{fig:2}(b), turning out to be in the range of (100$\pm$50)~G$\Omega$, when excluding the offset-afflicted values in the case of the 2-3 diagonal. These diagonal DW resistances appear to be two orders of magnitude lower as compared to both the T2T-DW and the bulk resistances that lie in the range of 10~T$\Omega$ (shown in figs.~\ref{fig:2}(a) and (b) as well). Note that the measured H2H DW resistances are comparable to the values reported earlier by Sluka et al.\cite{slu13}.

\begin{figure}[!ht]
\includegraphics[width=0.5\textwidth]{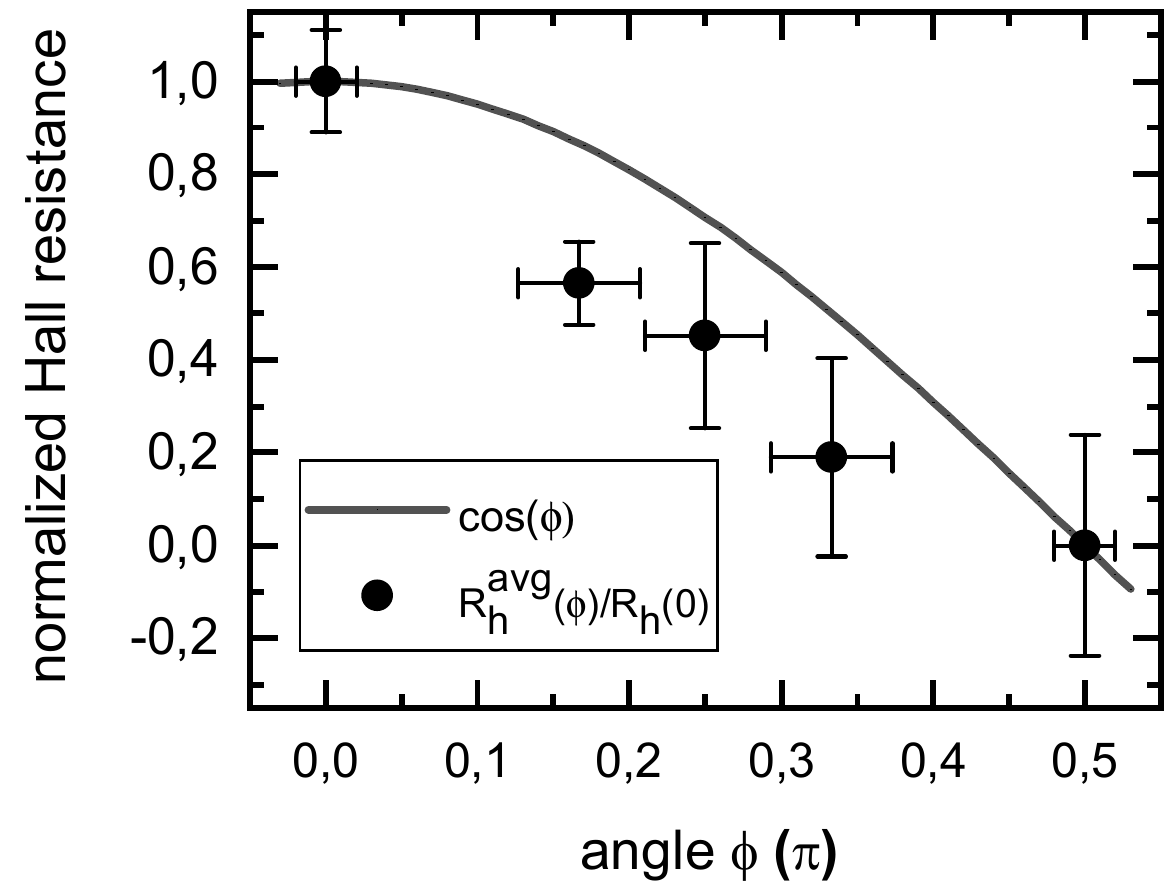}
\caption{\label{fig:angle_dep_main}Angular dependence of the averaged normalized Hall resistance measured with a constant absolute value of the magnetic field of 400~mT. The DW normal is rotated relative to the $B$-field vector by the angle $\phi$ [see sketch in SI-fig.~S2(b)]. In turn, the normalized Hall resistance shows nearly the cosine function, which is indeed expected from the angle dependence of the Lorentz force, thus giving strong motivation to proceed with in-depth Hall-effect investigations. Note that the $R_h$ values were averaged over both B-field- and current directions, respectively, the raw data before B-field averaging is depicted in SI-fig.~S2(a).}
\end{figure}

\begin{figure*}[!ht]
\includegraphics[width=\textwidth]{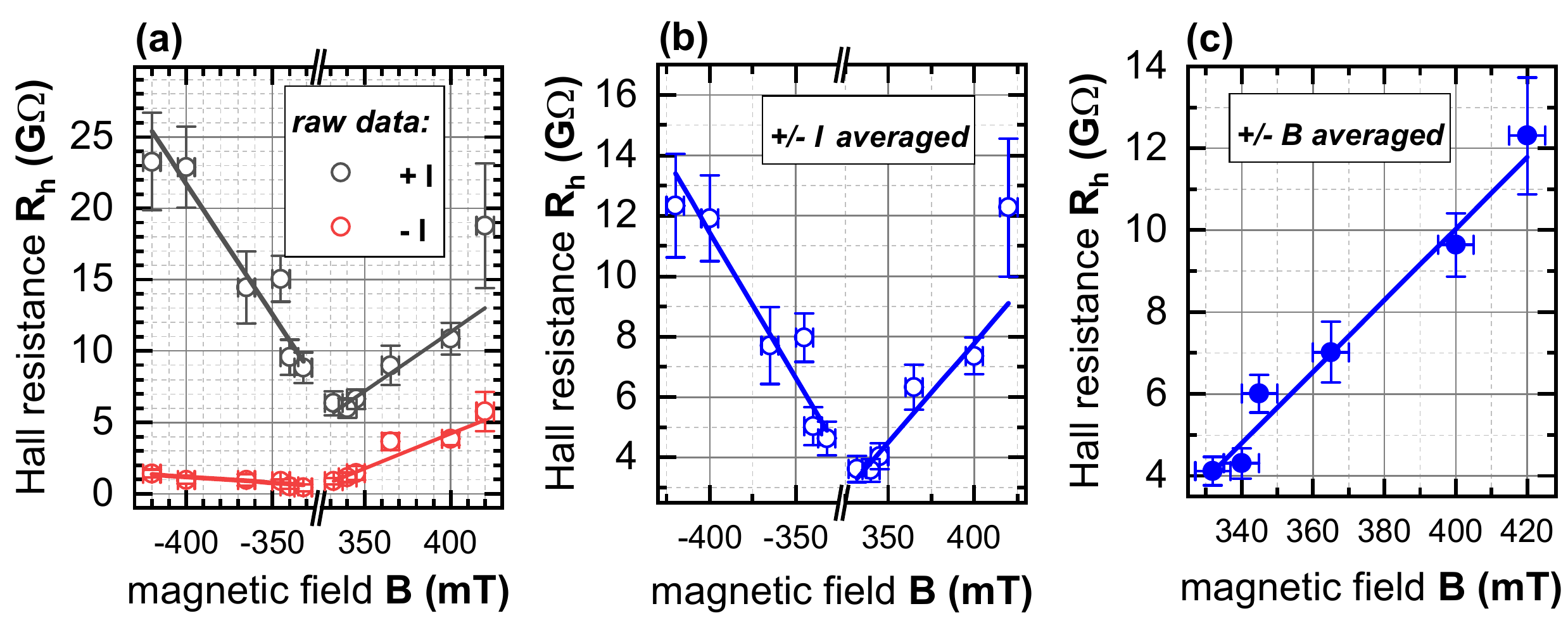}
\caption{\label{fig:4}Results of macroscopic Hall-effect measurements at a maximally charged 90$\symbol{23}$~[110] BaTiO$_3$ domain wall -- Hall resistance after several stages of data processing: (a) Raw Hall resistance ($R_h$) values for six different B-field strengths and both signs of B-field- and measuring current directions. (b) Hall resistance as a function of the magnetic-field strength after averaging over both current directions. (c) Linear $R_h$-vs.-$B$ dependence after averaging over both magnetic-field directions. The slope of the fit curve is reciprocal to the product $n_{2D} \cdot q$, with the elementary charge $q$ [eqs.~(\ref{eq:2D_carrier_density}) and (\ref{eq:offset_contributions})], which allows to extract the 2D charge carrier density, reading $n_{2D}\approx$7$\times$10$^3$cm$^{-2}$.}
\end{figure*}

Proceeding with the key experiment of the present study, i.e., the quantification of the Hall voltage within the conductive DW, the BTO crystal was placed within the magnetic field of an electromagnet providing fields of up to 420~mT [fig.~\ref{fig:1}(d)]. Contacts 1 and 4 were connected to a Keithley 6517B electrometer providing a bias voltage of $\pm$40~V, which resulted in currents $I=I_{14}$ in the range of typically 0.1~nA flowing along the DW. The corresponding charge carriers that experience a Lorentz force $F_L$ as sketched in fig.~\ref{fig:1}(d) provoke the Hall voltage $U_H:=U_{23}$, which is detected between contacts 2 and 3 using a Keithley 2700 multimeter. 

As a first rough test for the existence of a Lorentz force (and thus any Hall effect at all), the angle dependence of a presumed Hall voltage, i.e., $U_{23}$, was determined by rotating the DW plane by distinct angles $\phi$ in a fixed magnetic field of 400~mT. Thereby, $\phi$ denotes the angle between the B-field vector and the normal of the DW plane [cf. sketch in~SI-fig.~S2(b)]. The ratio between the Hall voltages at $\phi$ and at zero degree is denoted as normalized Hall voltage and should theoretically follow a cosine function due to the vector product nature of the Lorentz force. As seen from fig.~\ref{fig:angle_dep_main} displaying the normalized Hall voltage after averaging over the two different current and magnetic-field directions [for the raw data before field-averaging cf. SI-fig.~S2(a)], we observe a good agreement with the above expectation. These results motivated us to continue with an in-depth Hall-effect analysis.

However, before we straightly continue with any further and more elaborate experimental results, let us first consider the Hall effect in its simplest form, i.e., assuming (i) the current being established by electrons as majority charge carriers (due to their superior mobility as compared to holes at the one hand and the positive bound charges at a H2H-DW requiring negative screening charges at the other hand \cite{bed15,wer17,xia18}), as well as (ii) the magnetic field pointing perpendicular to the conducting layer (the DW). Then, the Hall voltage $U_H$ follows as \cite{Schroder2006,lun09}:
\begin{equation}
\label{eq:Hall_voltage}
    U_H=\frac{r\cdot I\cdot B}{q\cdot n\cdot d} \quad ,
\end{equation}
with $B$ being the absolute value of the magnetic field, $I$ the current driven through the conducting layer (here: $I=I_{14}$), $q$ the elementary charge, $d$ the layer thickness in magnetic-field direction (here: the DW thickness), $n$ the three-dimensional charge carrier density, while $r$ stands for the so-called Hall factor, which depends on the internal scattering mechanisms and lies typically between 0.5 and 2. Since the latter is not exactly known here, it is -- as a common practice -- assumed to be unity\cite{cam16}, which means that the derived mobilities and carrier densities are "Hall-specific" quantities, which differ from the respective "actual" quantities by a factor of $r$. Introducing the two-dimensional charge carrier density $n_{2D}$ with $n_{2D}=n\cdot d$ and using $R_h=U_H/I=U_{23}/I_{14}$ denoted as the Hall resistance (not to be confused with the Hall coefficient $R_H=\pm r/(q\cdot n)$, which is often used in the literature), $n_{2D}$ can be extracted from experimentally accessible quantities, i.e., $I_{14}$, $U_{23}$, $B$, via:
\begin{equation}
\label{eq:2D_carrier_density}
    n_{2D}=\frac{B}{q\cdot R_h} \quad .
\end{equation}
For the calculation of the Hall mobility $\mu$ of the majority charge carriers, we employ the relation:
\begin{equation}
\label{eq:Hall_mobility}
    \mu=\frac{1}{q\cdot n_{2D}\cdot R_s} \quad .
\end{equation}
There, $R_s$ is the sheet resistance of the DW, which is obtained by solving van der Pauw's equation numerically\cite{lun09}:
\begin{equation}
\label{eq:vdP_sheet_resistance}
    \exp{\left (-\pi\frac{R_{13,42}}{R_s}\right )}
    +\exp{\left (-\pi\frac{R_{34,21}}{R_s}\right )}
    =1 \quad .
\end{equation}
For that, $R_{13,42}=U_{42}/I_{13}$ and $R_{34,21}=U_{21}/I_{34}$ have to be determined from corresponding current injection and voltage measurements between the respective contacts [cf.~fig.~\ref{fig:1}(d)]; the measured values are listed in the SI-table~S1, and the graphical solution of eq.~(\ref{eq:vdP_sheet_resistance}) is illustrated in SI-fig.~S3. We note that neither for the subsequent evaluation of $n_{2D}$ nor for the calculation of $\mu$ the exact knowledge of the DW thickness is needed! 

Summarizing the outcome of the previous four equations, it is obvious that already the experimental determination of $R_s$ (at zero field) and $R_h$ (at a fixed $B$-field) allows us to extract $n_{2D}$ from eq.~(\ref{eq:2D_carrier_density}) and $\mu$ from eq.~(\ref{eq:Hall_mobility}) easily. 

However, the applicability of eq.~(\ref{eq:Hall_voltage}) and, in particular eq.~(\ref{eq:vdP_sheet_resistance}), requires several restrictive assumptions: the conducting sheet in the van-der-Pauw configuration needs to be homogeneous, isotropic, uniform in thickness, without holes, while the contacts need to be point contacts at the perimeter\cite{vdP58}, and the temperature must be perfectly constant at any point of the current path. These preconditions are rarely achieved in real measurement scenarios. Thus, the approach is susceptible to rather large errors. Especially, the actually measured voltage between contacts 2 and 3, denoted as $U_m$ for the moment, contains several offset contributions, as, f.i., comprehensively explained in the review article by Werner\cite{wer17b}, exhibiting the following general form:
\begin{equation}
\label{eq:offset_contributions}
    U_m=I \cdot R_s(\mu \cdot B + \alpha) + U_0 \quad .
\end{equation}
There, $U_0$ represents a current-independent thermal offset voltage, which can become significant already for small temperature differences between the contacts, while $\alpha$ is a geometrical factor originating from non-perfect sample and contact symmetries. The latter effect leads to a current-dependent misalignment voltage $\alpha \cdot I \cdot R_s$. Measuring $U_m$ under forward and reverse current, makes it possible to get rid of that constant thermal offset according to $U^{\ast}=\frac{1}{2}[U_m(+I)-U_m(-I)]$, while the misalignment problem is typically circumvented by acquiring $U^{\ast}$ at two opposite B-fields yielding:
$U_H=\frac{1}{2}[U^{\ast}(+B)-U^{\ast}(-B)]$. However, an even more
elaborate correction scheme, as discussed by Werner\cite{wer17b}, can be realized by sweeping the magnetic field, in order to fit $U_H(B)$, or analogously, $R_h(B)$ linearly and extract $n_{2D}$ from the slope $b=(q \cdot n_{2D})^{-1}$ according to a repositioned eq.~(\ref{eq:2D_carrier_density}).

Taking the above theoretical considerations into account, we acquired the voltage $U_{23}$ at six different B-field strengths between 330~mT and 420~mT for both directions of the measuring current and the magnetic field, respectively, resulting in four times six points of raw data. The respective Hall resistances ($R_h=U_m/I_{14}=U_{23}/I_{14}$ in this case) are plotted in fig.~\ref{fig:4}(a), where any combination of current and field directions shows different slopes and offsets, while averaging over the two current directions, as visible in fig.~\ref{fig:4}(b), in a first step, leads to already fairly similar slopes. Subsequently, an averaging over the two B-field directions is realized, leading to the final -- within the error bars -- satisfying linear $R_h$-vs.-$B$ dependence depicted in fig.~\ref{fig:4}(c). 

From the slope of this characteristic curve, a 2-dimensional charge carrier density $n_{2D}$ of (7.2$\pm$1.4)$\times$10$^3$~cm$^{-2}$ can be derived [eq.~(\ref{eq:2D_carrier_density})]. With the determination of the sheet resistance $R_s$ according to eq.~(\ref{eq:vdP_sheet_resistance}), which turned out to be 2.1~T$\Omega$ (see SI-tab.~S1 and SI-fig.~S3 for details) the Hall mobility $\mu$ is calculated as (395$\pm$81)~cm$^{2}$(Vs)$^{-1}$ employing eq.~(\ref{eq:Hall_mobility}). In comparison to mobilities of BaTiO$_3$ bulk crystals, which have been reported\cite{ber67,yoo04} to be in the range of 1~cm$^2$(Vs)$^{-1}$ at room temperature, the DW Hall mobility found here is significantly larger and in the range observed for the ferroelectric YbMnO$_3$ and ErMnO$_3$ domain walls\cite{cam16,tur18}, and also comparable to values observed in different topical 2-dimensional electron systems such as MoS$_2$ \cite{rai18}, which is a very promising result. At this point, we emphasize that extracting $n_{2D}$ and $\mu$ via the slope from the B-field dependence of the Hall resistance (and not from obtaining only single, even offset-corrected values of $R_h$ under a constant B-field) is mandatory in order to prevent errors of several 100\%, as has been illustrated in detail in SI-fig.~S5. The problem becomes obviously less significant for larger B-fields of several Tesla, which were, however, not available in the present setting. 

Finally, we provide an estimation of the 3-dimensional Hall charge carrier density $n$. By assuming the charged-domain-wall width $d$ to be in the range\cite{slu12,slu13} of $10-100$~nm, an interval for this quantity can be derived reading $n$ between 0.7$\times$10$^9$ and 7.2$\times$10$^9$~cm$^{-3}$. If domain wall widths as thin as 10 to 100~pm, as derived from transmission electron microscopy analyses\cite{gon16} -- however, observed for neutral LiNbO$_3$ DWs exhibiting sharp DWs down to seven unit cells -- are assumed as the lower limit for $d$, the corresponding 3D charge densities would be by a factor of 1000 higher, i.e., in the 10$^{12}$cm$^{-3}$ range. However, this issue remains speculative and the quantitative values for $\mu$ and $n_{2D}$ are the virtually robust output of the present study.

 In summary, an elaborate approach to quantify electrical transport parameters of conducting ferroic domain walls that penetrate the whole crystal volume, based on Hall-effect measurements has been demonstrated. In particular, the well-established van-der-Pauw technique was adopted in the way that domain walls were electrically contacted by four vapor-deposited electrodes, two on each crystal surface. The Hall resistance $R_h$ as a function of the magnetic field $B$ was recorded exemplarily as a proof-of concept in a straight highly conductive 90$\symbol{23}$- head-to-head domain wall in (110) BaTiO$_3$. The charge carrier (electron) density in two dimensions $n_{2D}$ was calculated from the slope of the offset-corrected $R_h$-vs.-$B$ characteristics. Moreover, the sheet resistance $R_s$ was determined according to van der Pauw's classical technique, and in turn the Hall mobility $\mu$ derived, which appeared to be competitive with other 2D electronic systems. After having shown the feasibility of the technique for this single model DW in BaTiO$_3$, the methodology should be extended to other model ferroelectric materials with stable CDWs such as is LiNbO$_3$. Notably, the present approach allows for correcting both temperature and size effects of the electrodes occurring herein.
 
 As an outlook beyond the present work, technical refinements employing (i) lithographically defined electrode structures, which potentially provide higher compliance with part of the van-der-Pauw restrictions on the one hand, and (ii) a further in-depth analysis of the whole resistivity tensor of the DW at zero field and in higher magnetic fields of several Tesla on the other hand, will be in focus for the future. Thereby, special attention should be laid on the the magnetoresistive (MR) response, since first test experiments showed a huge MR in 1--4 direction (SI-fig.~S4) and a similar strong MR of the sheet resistance $R_s$ would mean a certain impact on the derivation of the Hall mobility.


\begin{acknowledgement}
\begin{small}
This work was financially support by the Deutsche Forschungsgemeinschaft (DFG) through the joint DFG--ANR project TOPELEC (EN~434/41-1 and ANR-18-CE92-0052-1), the CRC~1415 (ID: 417590517), the FOR~5044 (ID:
426703838, \url{http://www.for5044.de}), as well as through the W\"urzburg-Dresden Cluster of Excellence on "Complexity and Topology in Quantum Matter" - ct.qmat (EXC 2147, ID: 39085490). Moreover, we acknowledge the support from the Czech Grant Agency (GACR project No. 20-05167Y) and from the Operational Program Research, Development and Education financed by the European Structural and Investment Funds, and from the Czech Ministry of Education, Youth and Sports (Project No. SOLID21-CZ.02.1.01/0.0/0.0/16\_019/0000760). The work was also supported by the Light Microscopy Facility of the CMCB Technology Platform at TU Dresden.
\end{small}
\end{acknowledgement}

\begin{small}

\begin{suppinfo}

For convenience, the Supporting Information is appended directly  at the end of the document in the arXiv preprint version, including a more detailed decsription of the material and the methodology, further electrical-characterization data, as well as considerations on error propagation. 

\end{suppinfo}
\end{small}
\begin{footnotesize}
\bibliography{BTO_DWC}
\end{footnotesize}


\beginsupplement

\onecolumn

\section*{Supporting Information}

\subsection{Materials and Methods}

\subsubsection{Crystal preparation} 

Ferroelectric H2H and T2T DWs were engineered in a (110)$_c$ oriented bulk BaTiO$_3$ single crystal grown with the top-seeded solution-growth method (supplied by EOT GmbH, \url{http://www.eotech-de.com}). Transparent (about 10~nm thick) golden electrodes were sputtered on (110) facets that allowed to observe and to control CDW formation. The used engineering technique, as described in detail in ref.~\cite{bed16}, is based on frustrative poling along the [110]$_c$ direction near the phase transition between orthorhombic and tetragonal phases. Free charge carriers compensating H2H and T2T DWs were generated by UV light illumination during poling. For the subsequent CSHG observations and Hall effect measurements on previously prepared H2H and T2T DWs, the golden electrodes were carefully removed using polishing diamond paste with 1-micron grains.

\subsubsection{Domain wall imaging} 

Domain wall imaging [cf.~figs.~1(a),(b)] was performed via confocal Cherenkov second harmonic generation (CSHG). A commercial setup was used consisting of a laser scanning microscope (SP5 MP, Leica), working in tandem with a tunable Ti:Saphire laser (Mai Tai BB, Spectra Physics). A 100-fs pulse at 900~nm was employed for scanning the sample with a galvanometric xyz-scanner. At the domain wall, the laser beam is converted into a CSHG-signal that then is recorded in back-reflection -- for more details refer to ref.~\cite{kir19}. The resulting image was taken at a frame rate of 0.1~Hz resulting in a 1024*1024 pixel image covering an area of 760~\textmu m*760~\textmu m.
The larger 2D overview scan (fig.~1(a) in the main text) was stitched together from multiple scans using the processing software packages \emph{ImageJ} and \emph{ParaView}.

\subsubsection{Current-voltage characterization}

For acquiring standard IU-characteristics between electrode pairs 1--4 and 2--3, respectively, the measuring voltage was swept from $-$50~V to $+$50~V in steps of 10~V using the voltage source of a Keithley 6517B electrometer with waiting intervals of 2~s while subsequently measuring the current with the same device. Each point was recorded 8 times by cycling up and down between $-$50~V to 50~V to rule out capacitive effects. The standard error of the resulting mean value was taken as the error bar. The T2T and bulk reference measurements were conducted ranging from $-$60~V to $+$60~V with a smaller step size of 6~V.

\subsubsection{Determination of the sheet resistance}

The sheet resistance was extracted by taking the active voltage source of a Keithley 6517B electrometer to apply $+$40~V between the contacts as requested by eq.~(4), while simultaneously tracking the resulting current for 60~s. A Keithley 2700 multimeter was used to record the resulting voltage between the remaining contacts for 240~s, with a sampling rate of 5~Hz.

\subsubsection{Hall-effect measurements}

The detection of the Hall voltage $U_{23}$ relies on the aforementioned Keithley 6517B electrometer as well. A measuring voltage of $\pm$40~V was applied between contacts 1 and 4 in order to drive the current $I_{14}$ that then was recorded by the same device, through the domain wall structure. A Keithley 2700 digital multimeter was utilized to measure the resulting Hall voltage between contacts 2 and 3. The measurement times and sampling rates were the same as used above when determining the sheet resistance.
An electromagnet, able to create magnetic fields ranging up to 420~mT, was used with the sample mounted onto a rotational table, allowing to cover an angular range between 0$\symbol{23}$--180$\symbol{23}$.
Note that the measurements were -- rather unusual for Hall effect experiments -- conducted by applying a voltage between two of the electrodes (1, 4), resulting in a current. A voltage source was used instead of a current source to ensure the samples' stability by not exceeding certain critical voltage thresholds that might initiate DW motion. The current direction was changed by changing the sign of the applied voltage.


\subsection{Complementary electrical-transport data}

\subsubsection{Linear IU-characteristics}

\begin{figure*}[!h]
\includegraphics[width=\textwidth]{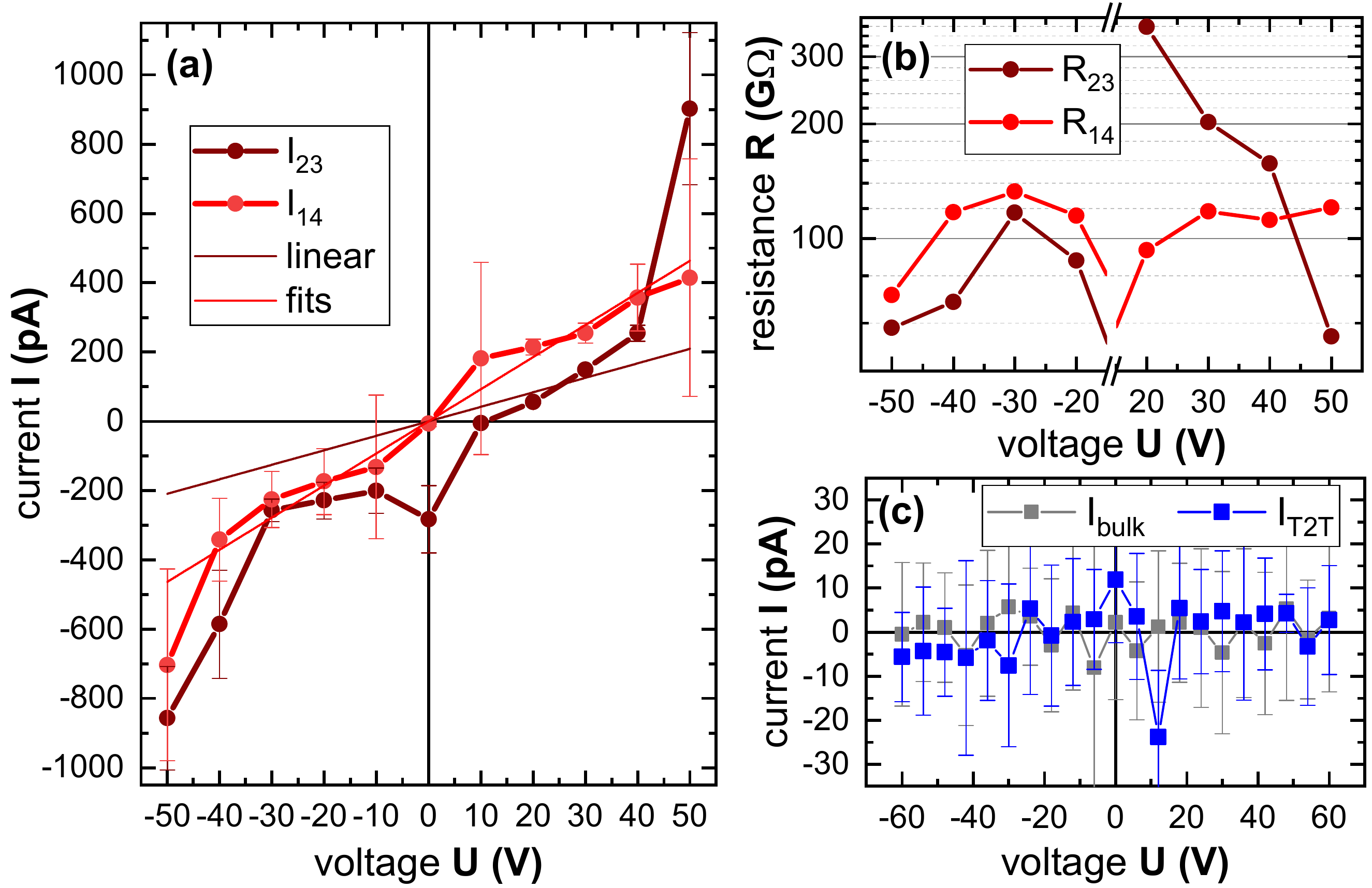}
\caption{\label{fig:IU_linear} (a) Linear representation of the current-voltage (IU) characteristics of the conductive H2H domain wall for both diagonal electrode configurations including linear fits. (b) Domain wall resistance values as extracted from the data of subfigure (a) neglecting the values around zero voltage. (c) Reference IU-curves for a non-conductive T2T domain wall and for the bulk of the (110) BaTiO$_3$ crystal under investigation. Note that fig.~2 of the main text contains the same data, however, plotted semilogarithmically.}
\end{figure*}

\clearpage

\subsubsection{Angle-dependent Hall resistance}

\begin{figure*}[!h]
\includegraphics[width=0.9\textwidth]{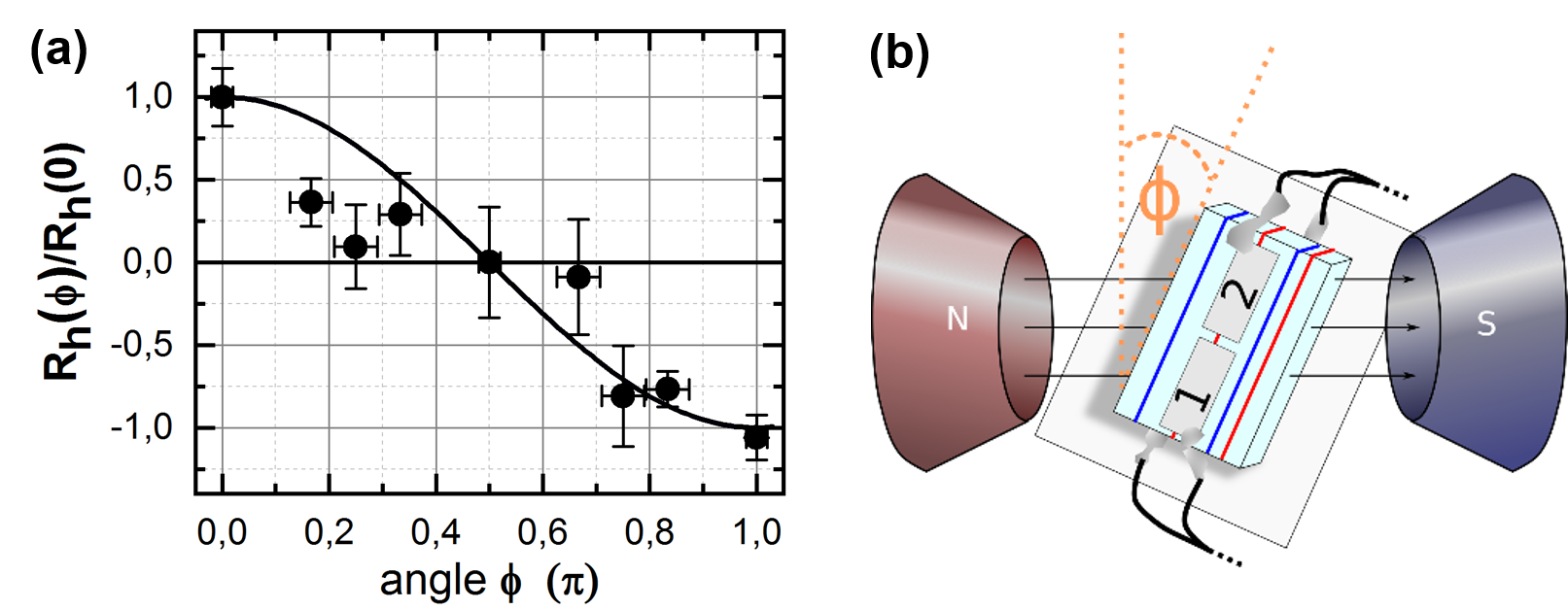}
\caption{\label{fig:angle_dep_raw}(a) Angular dependence of the normalized Hall resistance of the conductive H2H DW -- raw data (dots) of the averaged representation shown in fig.~3 of the main text, plotted together with the expected cosine function (solid line). (b) Sketch of the measurement geometry.}
\end{figure*}


\subsubsection{Sheet resistance}

\begin{table}[!h]
\caption{\label{tab:Raw_Data_sheet_resistance}%
Current and voltage values recorded in the van-der-Pauw configuration, as well as the directly derived resistance values $R_{13,42}$ and $R_{34,21}$. The latter two serve as input when extracting the sheet resistance $R_s$ according to eq.~(4) in the main text.}
\renewcommand{\arraystretch}{1.5}
\begin{tabular}{cccccc}
\hline
  $R_{13,42}$ (10$^{12}\Omega$) & $U_{42}$ (V) & $I_{13}$ (10$^{-12}$A) & $R_{34,21}$ (10$^{12}\Omega$) & $U_{21}$ (V) & $I_{34}$ (10$^{-12}$A)\\
\hline
0.9 & 0.6 & 0.7 & 0.2 & 0.4 & 2.2 \\
\hline
\end{tabular}
\end{table}

\begin{figure}[!h]
\includegraphics[width=0.45\textwidth]{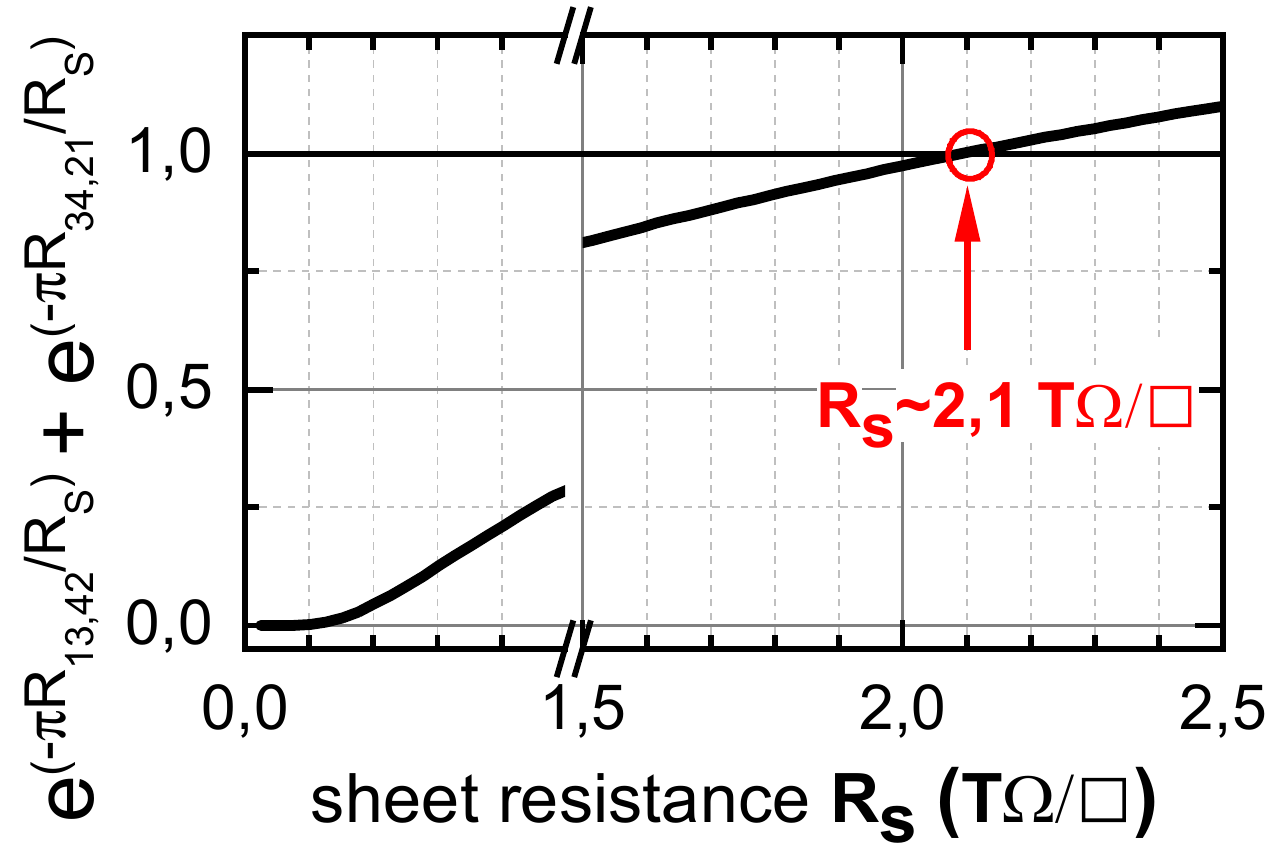}
\caption{\label{fig:vdP_sheet_resistances}Determination of the H2H domain walls' sheet resistances $R_s$ from graphical solution of eq.~(4) employing the measured resistance values $R_{13,42}$ and $R_{34,21}$ (cf.~SI-tab.~\ref{tab:Raw_Data_sheet_resistance}).}
\end{figure}

\clearpage

\subsubsection{Diagonal magnetoresistance}

\begin{figure}[!h]
\includegraphics[width=0.4\textwidth]{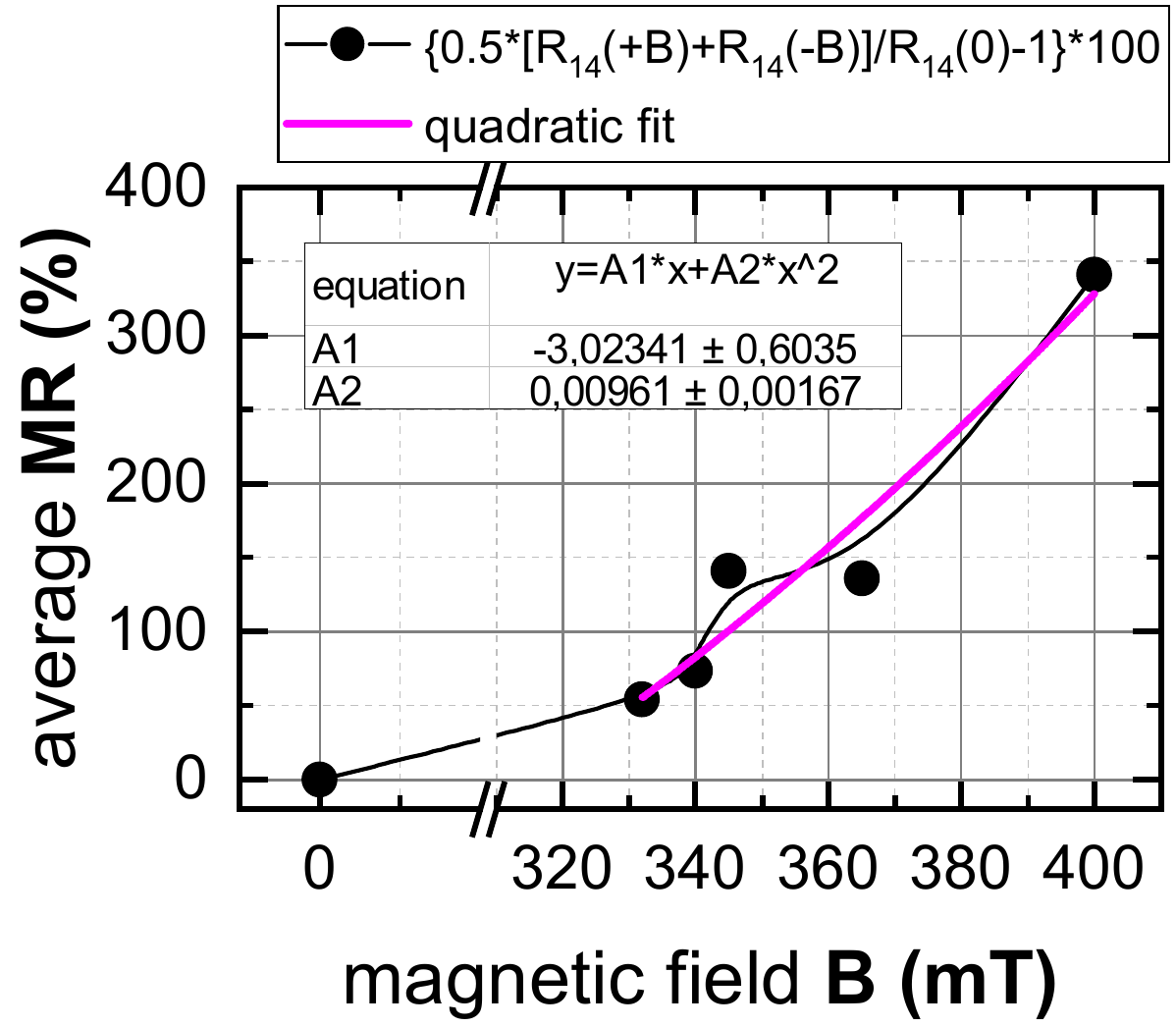}
\caption{\label{fig:MR}Exemplary measurement of the magnetoresistance, employing the definition MR$:=[R(B)/R(0)-1]\cdot 100\%$, along the path between the electrodes 1 and 4 in the conductive H2H DW.}
\end{figure}


\subsubsection{Considerations on error progagation}

\begin{figure*}[!h]
\includegraphics[width=0.75\textwidth]{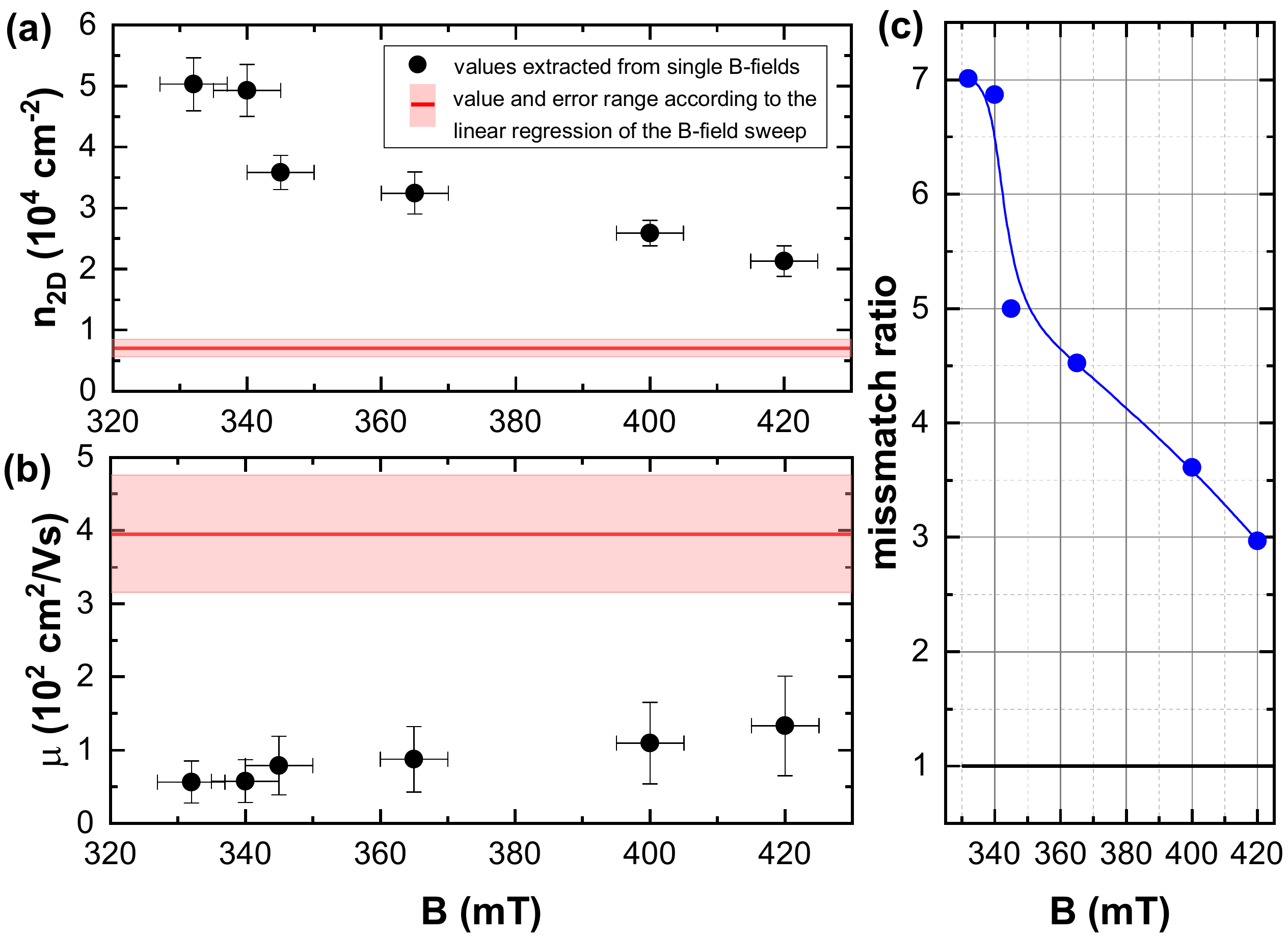}
\caption{\label{fig:errors}Discrepancies of the numerical values for the (a) 2D electron density $n_{2D}$ and (b) Hall mobility $\mu$ derived from Hall effect measurements at single magnetic fields (squares) and from the slope of the $B$-field sweep (red range). To illustrate the size of the error made when using only single B-fields instead of the slope of the $R_h$-vs.-$B$ characteristics, the ratios between the two results are depicted in panel (c) showing a huge discrepancy by a factor of 7 (i.e., 700\% deviation!) for lower magnetic fields with a strongly decreasing tendency towards higher fields.}
\end{figure*}

\end{document}